\documentclass[reprint, aip, apl]{revtex4-1}

\usepackage{graphicx}
\usepackage{xcolor}
\usepackage[colorlinks=true, linkcolor=blue, citecolor=blue, urlcolor=blue]{hyperref}
\usepackage{units}

\begin{document}

\title{Ambipolar quantum dots in undoped silicon fin field-effect transistors}

\author{Andreas V. Kuhlmann}
%\email{ndr@zurich.ibm.com}
\affiliation{IBM Research-Z\"urich, S\"aumerstrasse 4, CH-8803 R\"uschlikon, Switzerland}

\author{Veeresh Deshpande}
\affiliation{IBM Research-Z\"urich, S\"aumerstrasse 4, CH-8803 R\"uschlikon, Switzerland}

\author{Leon C. Camenzind}
\affiliation{Department of Physics, University of Basel, Klingelbergstrasse 82, CH-4056 Basel, Switzerland}

\author{Dominik M. Zumb\"uhl}
\affiliation{Department of Physics, University of Basel, Klingelbergstrasse 82, CH-4056 Basel, Switzerland}

\author{Andreas Fuhrer}
\affiliation{IBM Research-Z\"urich, S\"aumerstrasse 4, CH-8803 R\"uschlikon, Switzerland}

\date{\today}

\begin{abstract}
We integrate ambipolar quantum dots in silicon fin field-effect transistors using exclusively standard complementary metal-oxide-semiconductor fabrication techniques. We realize ambipolarity by replacing conventional highly-doped source and drain electrodes by a metallic nickel silicide with Fermi level close to the silicon mid-gap position. Such devices operate in a dual mode, either as classical field-effect or single-electron transistor. We implement a classical logic NOT gate at low temperature by tuning two interconnected transistors into opposite polarities. In the quantum regime, we demonstrate stable quantum dot operation in the few charge carrier Coulomb blockade regime for both electrons and holes.
\end{abstract}

\maketitle
Quantum information can be encoded in the spin state of a single electron or hole confined to a semiconductor quantum dot (QD)~\cite{Loss1998,Kloeffel2013,Hanson2007}. Several material systems have been explored in the search of a highly coherent spin quantum bit (qubit). Silicon (Si) is a particularly promising material platform for scalable spin-based quantum computing because of its fully developed, industrial manufacturing processes, which enable reliable and reproducible fabrication at the nanometer scale~\cite{Zwanenburg2013,Veldhorst2014,Maurand2016}. Furthermore, natural silicon consists of \unit[95]{\%} non-magnetic nuclei (\unit[92]{\%} $^{28}$Si, \unit[3]{\%} $^{30}$Si), suppressing hyperfine-induced decoherence~\cite{Elzerman2004,Petta2005,Koppens2006}. A nearly nuclear-spin-free environment can additionally be engineered by means of isotopic purification~\cite{Itoh2003}. Electron spins in silicon are also subject to a weak spin-orbit interaction (SOI) and can thus be almost completely isolated from environmental noise~\cite{Kuhlmann2013}. As a result, an excellent dephasing time $T_2^{\mathrm{*}}$ of \unit[120]{$\mu$s} has been demonstrated for the electron spin qubit in isotopically enriched silicon ($\geq$\,\unit[99.9]{\%} of $^{28}$Si)~\cite{Veldhorst2014}.

For scalable quantum circuits, qubit control via electric rather than magnetic fields is more promising in terms of speed and hardware implementation. In this regard, the hole spin represents an attractive alternative to its electron counterpart~\cite{Fischer2008,Brunner2009,Prechtel2016}. 
The asymmetry of the silicon band structure with respect to the conduction (CB) and valence bands (VB) manifests itself in different characteristics for electrons and holes. While the electron Bloch function has s-wave symmetry, the hole has p-wave symmetry. Consequently, hole spins experience a weaker hyperfine, yet stronger SOI, which enables fast, all-electrical spin manipulation~\cite{Prechtel2015,Crippa2018,Voisin2016}. Despite these potential benefits, hole spin qubits in silicon are still largely unexplored. Recently, qubit functionality with fast, purely electrical, two-axis control was shown for a hole spin, yet with inferior coherence compared to the electron spin~\cite{Maurand2016}. 

Usually, either electrons~\cite{Angus2007,Lim2009,Lim2009a,Prati2012} or  holes~\cite{Li2013,Li2015,Maurand2016,Voisin2016,Liles2018} are confined in silicon QDs. Ambipolar devices, by contrast, can be operated in both the electron and hole regime~\cite{Martel2001,Jarillo-Herrero2004,Byon2007,Colli2007,Guettinger2009,Chen2012,Li2006,Ishikuro1999}. For planar silicon metal-oxide-semiconductor (MOS) QD structures, ambipolar behavior was demonstrated by integrating both \emph{n}- and \emph{p}-type reservoirs on the same device~\cite{Betz2014,Mueller2015a,Mueller2015b,Spruijtenburg2016,Spruijtenburg2018}. Ambipolar devices provide great flexibility for scalable spin-based quantum circuits, since both types of charge carriers can be manipulated in exactly the same crystalline environment, allowing for direct benchmarking of hole against electron spin qubits. 
\begin{figure}[b]
\centering \includegraphics[width=\linewidth]{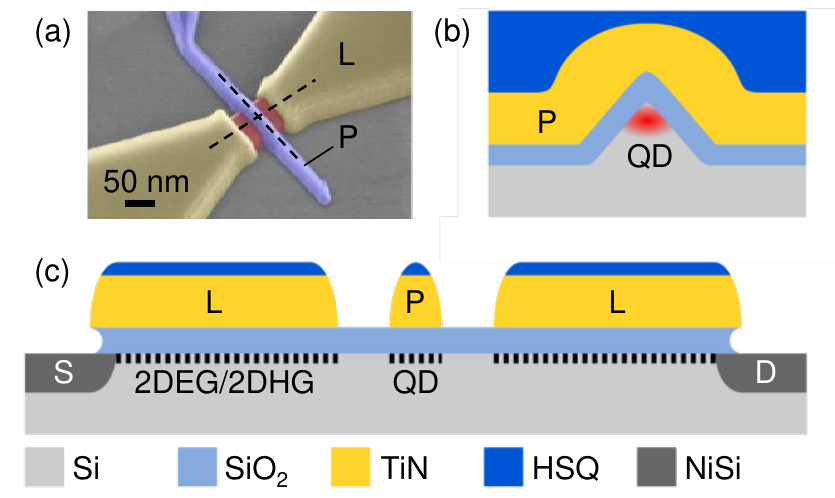}
\caption{Device layout: (a) False-color scanning electron micrograph showing a FinFET structure before sealing it with a SiO$_2$ passivation layer. Devices consist of a single plunger (blueish color) and two lead gates (yellowish color) to silicided source and drain. The gate electrodes are operated in accumulation mode and are wrapped around the silicon fin (reddish color) defining the transistor channel. The dashed lines indicate the orientation of the schematic cross sections perpendicular to (b) and along (c) the fin. For a positive (negative) gate voltage electrons (holes) are accumulated at the Si-SiO$_2$ interface below the electrode. While the lead gates are designed to induce a two-dimensional electron (hole) gas opening low-resistance leads to source and drain, the plunger gate allows for local electrostatic control of the channel and to create a QD, which is located at the apex of the triangular shaped fin.}
\label{fig1}
\end{figure}

Here, we report on a new generation of ambipolar silicon QD devices based on today's industry standard, non-planar fin field-effect transistors (FinFETs)~\cite{Intel2011,Auth2012,Auth2017}. In an overlapping-gate structure, ambipolarity is achieved by using a metallic nickel silicide (NiSi) with Fermi energy close to the silicon mid-gap for source (S) and drain (D) electrodes~\cite{Larson2006,Colli2007,Byon2007}. This approach allows for a highly compact device layout, is easy to integrate and fully compatible with complementary metal-oxide-semiconductor (CMOS) technology. We successfully operate the devices both in a classical and quantum mode~\cite{Roche2012}, demonstrating simple co-integration between silicon-based qubits and traditional CMOS control hardware. 
%In the quantum regime, stable operation in the few charge carrier Coulomb blockade regime is observed. 

The layout of the home-built devices is shown in \mbox{Fig.\ \ref{fig1}}. First, the fin structures are defined on a near-intrinsic silicon substrate ($\rho>\unit[5000]{\Omega\mathrm{cm}}$, (100) surface) by means of electron-beam lithography (EBL) and dry etching, yielding a fin height of $\simeq$\,\unit[25]{nm}. A sacrificial thermally grown silicon dioxide (SiO$_2$) layer, which is removed in buffered hydrofluoric acid, allows for narrowing of the fin width ($\gtrsim$\,\unit[10]{nm}) and cleaning of etch-induced surface damage. This procedure leads to an almost triangular cross section for the narrowest fins. Subsequently, the gate stack is deposited, consisting of a high-quality, thermally grown SiO$_2$ layer ($\sim$\,\unit[10]{nm}, breakdown voltage $\sim$\,\unit[10]{V}), covered by \unit[40]{nm} of titanium nitride (TiN). An uniform layer of TiN, which is wrapped around the silicon channel, is obtained by atomic layer deposition. The gate layer is patterned by means of EBL and dry etching of TiN, resulting in a gate length of $\gtrsim$\,\unit[25]{nm} at a gate-to-gate separation of $\gtrsim$\,\unit[50]{nm}. Conventional impurity-doped source and drain electrodes are replaced by a metallic, non-magnetic NiSi, forming a Schottky barrier at the S/D-to-substrate junction~\cite{Larson2006,Zwanenburg2009}. By choosing a mid-gap silicide, ambipolar operation is realized in a simple, highly compact design, as no complementary charge reservoirs are required. So far, ambipolar silicon QDs have only been implemented by integrating separate \emph{n}- and \textit{p}-type contacts to the same channel, enlarging the device's footprint~\cite{Betz2014,Mueller2015a,Mueller2015b,Spruijtenburg2016,Spruijtenburg2018}. The NiSi electrodes are formed by EBL, Ni evaporation, lift-off and low-temperature silicidation annealing at 475$^{\circ}$\,C for \unit[30]{min} in an argon ambient. Lateral Ni diffusion below the gates allows for tuning of the Schottky barrier width, and ensures that source and drain contacts operate in an ohmic regime. After silicidation, unreacted Ni is selectively removed in order to avoid any magnetic impurities to be present in the device. Finally, the devices are protected from contamination by a SiO$_2$ passivation layer and are accessed via tungsten interconnects.   

The data presented here is obtained from direct current electrical transport measurements with the sample cooled to $T \simeq$ \unit[1.5]{K}. The devices' gate layer consists of a central plunger (P) and individual lead (L) gates to source and drain electrodes, as shown in Fig.\ \ref{fig1} (c). The gates are operated in accumulation mode: for a negative (positive) applied voltage holes (electrons) are accumulated at the Si-SiO$_2$ interface. Therefore, a two-dimensional electron (2DEG) or hole gas (2DHG) forms beneath the lead gates, acting as electrostatically defined source and drain, while the plunger gate induces a Coulomb island that defines the QD. The gaps separating lead and plunger gates create tunnel barriers between them~\cite{Hofheinz2006}.

\begin{figure}
\centering \includegraphics[width=\linewidth]{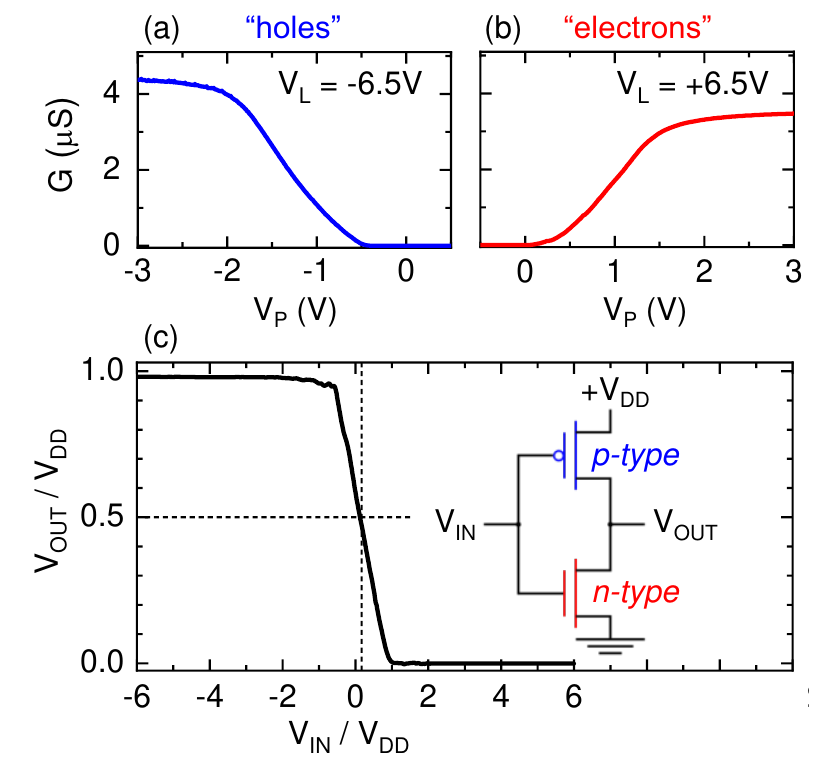}
\caption{Ambipolar turn-on curves and CMOS logic at cryogenic temperatures: conductance G versus plunger gate voltage V$_\mathrm{P}$ for the hole (a) and electron (b) regime. The lead voltage V$_{\mathrm{L}}$ is kept fixed at a value well above threshold, and the source-drain voltage V$_{\mathrm{SD}}$ is -1 V for holes and +1 V for electrons, where the bias polarity is chosen such that electrical stress on the device is minimized. (c) Voltage transfer characteristics and the corresponding circuit diagram of a CMOS inverter, consisting of a p-type and a complementary n-type FinFET. For \emph{low} input V$_{\mathrm{IN}}<0$ V the output V$_{\mathrm{OUT}}$ is \emph{high} and vice versa. All the measurements are performed at $T=1.5$~K.}
\label{fig2}
\end{figure}
\begin{figure*}[t]
\centering \includegraphics[width=\linewidth]{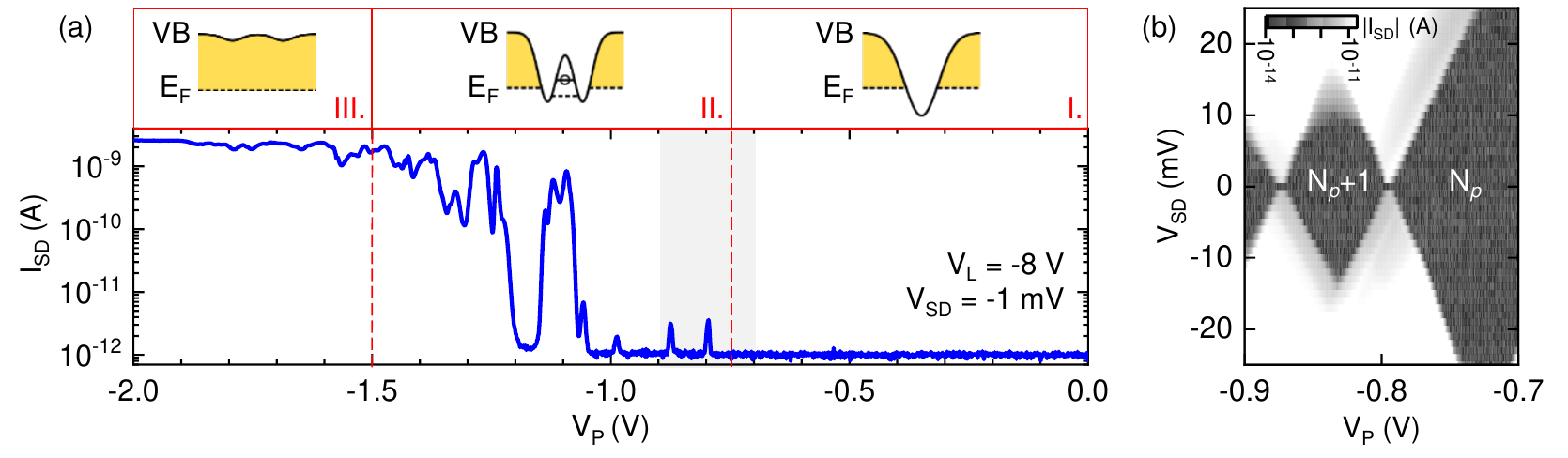}
\caption{QD formation: (a) source-drain current I$_{\mathrm{SD}}$ versus plunger gate voltage V$_{\mathrm{P}}$ in the low-bias regime (V$_{\mathrm{SD}}$ = -1 mV) on the hole side (V$_{\mathrm{L}}$ = -8 V). Top panel: Sketches of the real space band alignment in the vicinity of the plunger gate for the observed three different conductance regimes. (b) Charge stability map of the first measurable Coulomb resonance. The corresponding plunger gate voltage range is indicated by the grey shaded area in (a). The Coulomb diamonds are labeled with the number of holes residing on the dot (N$_p$ $\gtrsim 0$). All the measurements are performed at $T=1.5$ K.}
\label{fig3}
\end{figure*} 
First, the devices are operated in a classical field-effect transistor (FET)-like regime. Ambipolar transistor turn-on curves, revealing both \emph{n}- and \emph{p}-type conduction, are presented in Figs.\ \ref{fig2} (a), (b). The linear conductance G is plotted versus plunger gate voltage V$_{\mathrm{P}}$ at a constant lead gate voltage V$_{\mathrm{L}}$ of $\pm$\,\unit[6.5]{V}, creating conducting channels beneath the lead gates. A large source-drain voltage V$_{\mathrm{SD}}$ of $\pm$\,\unit[1]{V} ensures that the current is not dominated by charge carrier tunneling processes. The measurement reveals a slight asymmetry in current-onset voltages with respect to zero for electrons and holes: for V$_{\mathrm{P}}\gtrsim$ \unit[0]{V} \emph{n}-type and for V$_{\mathrm{P}}\lesssim$ \unit[-0.35]{V} \emph{p}-type conduction occurs. In between, the Fermi level lies in the band gap of silicon and no states are available for transport. This asymmetry is not fully understood and most likely a combination of various effects, such as a residual wafer background doping, charge traps or the metal gate work function~\cite{Mueller2015a,Spruijtenburg2016,Spruijtenburg2018}. The lower saturation current for electrons may also be due to an asymmetry of the silicide Schottky barrier for electrons and holes.

Any CMOS circuit can in principle be constructed using ambipolar transistors as sole building blocks. In the inset of Fig.\ \ref{fig2} (c) the most basic logic circuit - the CMOS inverter - is shown schematically. It consists of two complementary transistors connected at the gate and drain terminals. The inverter output voltage V$_{\mathrm{OUT}}$ is taken from the common drain electrode and is limited to the supply voltage V$_{\mathrm{DD}}$, which is applied to the \emph{p}-type transistor's source contact. The voltage transfer curve of our home-built inverter is presented in Fig.\ \ref{fig2} (c).  As the input voltage V$_{\mathrm{IN}}$ is varied from \emph{low} to \emph{high}, the inverted input signal is measured at the output, going from \emph{high} to \emph{low}. The \emph{high} output level gets with 0.98\,V$_{\mathrm{DD}}$ close to the ideal limit. The transition zone, however, is centered around V$_{\mathrm{IN}}\simeq 0.15$\,V$_{\mathrm{DD}}$ and not V$_{\mathrm{DD}}/2$ as the devices are not perfectly matched in threshold voltage and amplification. Nevertheless, the successful operation of an inverter at low-temperature proves that classical CMOS logic can be performed.

\begin{figure}
\centering \includegraphics[width=\linewidth]{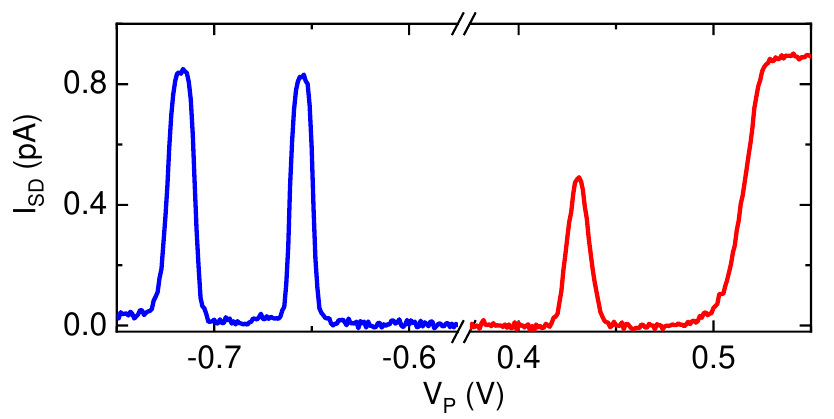}
\caption{Ambipolar Coulomb blockade: source-drain current I$_{\mathrm{SD}}$ versus plunger gate voltage V$_{\mathrm{P}}$ for holes (blue, V$_{\mathrm{L}}$ = -4 V, V$_{\mathrm{SD}}$ = +0.5 mV) and electrons (red, V$_{\mathrm{L}}$ = +3.2 V, V$_{\mathrm{SD}}$ = +2.5 mV). The measurements are performed at $T=1.5$ K.}
\label{fig4}
\end{figure}

In the linear transport regime at V$_{\mathrm{SD}}=\unit[-1]{\mathrm{mV}}$, quantum confinement in an island, which forms below the plunger gate, gives rise to pronounced Coulomb oscillations. The plunger gate-dependent source-drain current I$_{\mathrm{SD}}$ at V$_{\mathrm{L}}=\unit[-8]{\mathrm{V}}$ is shown in Fig.\ \ref{fig3} (a). In this regime the current is dominated by hole tunneling. The measurement is performed on a device with a fin width of $\simeq$\,\unit[20]{nm}, plunger gate length of $\simeq$\,\unit[25]{nm} and plunger-to-lead-gate separation of $\simeq$\,\unit[25]{nm}. Three different regimes of hole transport, which are depicted schematically in the top panel of Fig.~\ref{fig3}~(a), are observed: (i) for V$_{\mathrm{P}}\gtrsim$ \unit[-0.8]{V} the barrier induced by the plunger gate prevents current flow. (ii) For V$_{\mathrm{P}}\lesssim$ \unit[-0.8]{V} a series of Coulomb resonances indicates single-hole tunneling via a QD formed beneath the plunger gate. In the valleys between the peaks the device operates in Coulomb blockade, i.e.\ the QD contains a fixed number of holes~\cite{Hanson2007,Zwanenburg2013}. As this number increases with more negative V$_{\mathrm{P}}$, the plunger gate's fringe fields lower the barriers and the QD starts to open (V$_{\mathrm{P}} \lesssim$ \unit[-1.05]{V}). (iii) For V$_{\mathrm{P}} \lesssim$ \unit[-1.5]{V} a conducting channel is opened and the current is limited by the series resistance of the device. Similar behavior is found for positive V$_{\mathrm{P}}$ on the electron side (see Fig.\ \ref{fig4}). 

The first measurable Coulomb resonances are investigated in more detail by means of bias spectroscopy. In Fig.\ \ref{fig3} (b) the charge stability diagram is shown for a plunger gate voltage range highlighted by the gray shaded area in Fig.\ \ref{fig3} (a). Within this range, the tunnel barriers are still well defined. Clear Coulomb diamonds with a fixed number of holes N$_p$ on the QD are observed. Outside the diamonds sequential tunneling of holes through the QD occurs. The small dimensions of the device and the closing of the Coulomb diamonds at zero bias suggest formation of a single QD. Moreover, similar coupling of the QD to both source and drain (from the shape of the diamond we determine that the source and drain lever arms differ by just $\sim$\,6\,\%) dictates a central location of the charge island below the plunger gate. The charging energy is determined to be $e^2/C_{\Sigma}\simeq$ \unit[16]{meV} that corresponds to a total capacitance $C_{\Sigma}$ of \unit[10]{aF}. The plunger gate voltage spacing of the Coulomb resonances yields a gate capacitance $C_{g}$ of \unit[2.1]{aF}. The latter is in good agreement with the calculated MOS plunger gate capacitance, which can be estimated by an equivalent planar capacitor $C_g = \epsilon_0 \epsilon_{\mathrm{SiO_2}}\mathrm{S}/t_{\mathrm{SiO_2}}\sim$\,\unit[3.5]{aF} with $\epsilon_{\mathrm{SiO_2}} = 3.9$ the dielectric constant, $t_{\mathrm{SiO_2}}$ the oxide thickness and S the surface area of the gate-fin overlap. The gate voltage lever arm is $\alpha=C_g/C_{\Sigma}\simeq0.21$. The large charging energy and the wide opening in V$_{\mathrm{SD}}$ of the last diamond could indicate that the device is operating in the single-hole regime. However, more sensitive charge detection methods and a device structure that offers more tunability are necessary to evaluate this~\cite{Hanson2007,Field1993}. The lines of increased conductance that run parallel to the diamond edges in Fig.\ \ref{fig3} (b) can be attributed to resonant tunneling processes~\cite{Escott2010}, for instance excited orbital states of the QD. Various devices have been measured, showing similar behavior and charging energies. However, instabilities and deviations from the ideal picture reveal that the device performance is affected by charge-trapping defects.

Ambipolar behavior in the low V$_{\mathrm{SD}}$ regime is demonstrated in Fig.\ \ref{fig4} where I$_{\mathrm{SD}}$ is plotted versus V$_{\mathrm{P}}$ for both the electron and hole regime. The data was measured on a different device with the same physical dimensions as the one of Fig.\ \ref{fig3} (the electron regime of this device suffers from charge traps). Both in the electron and hole transport regime Coulomb oscillations occur. However, the asymmetry in the band structure of silicon with respect to the conduction and valence bands manifests itself in asymmetric electrical transport characteristics for holes and electrons. While for electrons a single current peak exists before the barriers vanish, the hole side exhibits a similar behavior to the previous device with several Coulomb oscillations.

In conclusion, we have introduced a novel type of ambipolar silicon QDs, integrated in today's industry standard, non-planar FinFETs. By making use of a mid-gap silicide, ambipolar devices are realized with the footprint of unipolar structures. We successfully operate these devices in a classical as well as quantum mode, thus demonstrate the compatibility of silicon-based quantum circuits with traditional CMOS control hardware. Future devices with even smaller physical dimensions, improved charge noise performance and a second gate layer for in-situ adjustment of the tunnel coupling will probably allow us to reliably access the single-electron (hole) regime. Such devices will enable direct benchmarking of electron against hole spin qubits. Moreover, an interconnected array of ambipolar QDs will offer a blank canvas for building custom, on-the-fly reconfigurable ``quantum CMOS'' circuits, which in analogy to classical CMOS, utilize both \emph{n}- and \emph{p}-type devices.

We acknowledge technical support in device fabrication from Ute Drechsler, Antonis Olziersky and Ralph Heller. This work was supported by NCCR QSIT and the Georg H.\ Endress foundation.

\end{document}